\documentclass[twocolumn, floatfix,prl,showpacs,amsmath,amssymb,superscriptaddress]{revtex4}

\usepackage{graphicx}
\usepackage{dcolumn}
\usepackage{amssymb}
\usepackage{amsmath}
\usepackage{color}
\usepackage{bm}
\usepackage{verbatim}
\begin{document}
 
\title{
 Disentangling the effects of spin-orbit and hyperfine interactions on spin blockade}
\author{S.~Nadj-Perge}
\affiliation{Kavli Institute of Nanoscience, Delft University of Technology, PO Box 5046, 2600 GA Delft, The Netherlands}
\author{S.M.~Frolov}
\affiliation{Kavli Institute of Nanoscience, Delft University of Technology, PO Box 5046, 2600 GA Delft, The Netherlands}
\author{J.W.W.~van Tilburg}
\affiliation{Kavli Institute of Nanoscience, Delft University of Technology, PO Box 5046, 2600 GA Delft, The Netherlands}
\author{J.~Danon}
\affiliation{Kavli Institute of Nanoscience, Delft University of Technology, PO Box 5046, 2600 GA Delft, The Netherlands}
\affiliation{Dahlem Center for Complex Quantum Systems, Freie Universit\"{a}t Berlin, Arnimallee 14, 14195 Berlin, Germany} 
\author{Yu.V.~Nazarov}
\affiliation{Kavli Institute of Nanoscience, Delft University of Technology, PO Box 5046, 2600 GA Delft, The Netherlands}
\author{R.~Algra}
\affiliation{Philips Research Laboratories Eindhoven, High Tech Campus 11, 5656AE Eindhoven, The Netherlands}
\author{E.P.A.M.~Bakkers}
\affiliation{Kavli Institute of Nanoscience, Delft University of Technology, PO Box 5046, 2600 GA Delft, The Netherlands}
\affiliation{Philips Research Laboratories Eindhoven, High Tech Campus 11, 5656AE Eindhoven, The Netherlands}
\author{L.P.~Kouwenhoven}
\affiliation{Kavli Institute of Nanoscience, Delft University of Technology, PO Box 5046, 2600 GA Delft, The Netherlands}
\date{\today}

\begin{abstract}
We have achieved the few-electron regime in InAs nanowire double quantum dots. Spin blockade is observed for the first two half-filled orbitals, where the transport cycle is interrupted by forbidden transitions between triplet and singlet states.
Partial lifting of spin blockade is explained by spin-orbit and hyperfine mechanisms that enable triplet to singlet transitions. The measurements over a wide range of interdot coupling and tunneling rates to the leads are well reproduced by a simple transport model. This allows us to separate and quantify the contributions of the spin-orbit and hyperfine interactions.
\end{abstract}

\pacs{73.63.Kv,72.25.-b}

\maketitle

Spins in semiconductor quantum dots are possible building blocks for quantum information processing \cite{losspra98}. The ultimate control of spin states is achieved in electrically defined single and double quantum dots \cite{hansonrmp07}. Many semiconductors that host such dots exhibit strong spin-orbit and hyperfine interactions. On the one hand, these interactions provide means of coherent spin control \cite{nowackscience07, folettinatphys09}. On the other hand, they mix spin states. In double quantum dots, mixing of singlet and triplet states weakens spin blockade \cite{koppensscience05,johnsonnature05,pfundprl07,pfundprb07,churchillnatphys09}, which is a crucial effect for spin qubit operation \cite{pettascience05, koppensnature06}. Spin mixing due to hyperfine interaction was studied in GaAs double quantum dots, where spin-orbit coupling is weak \cite{koppensscience05, johnsonnature05, jouravlevprl06}. In InAs, besides the hyperfine interaction, also spin-orbit interaction has a considerable effect on spin blockade. Previous measurement on many electron double dots in InAs nanowires demonstrated that spin blockade is lifted by both interactions \cite{pfundprl07,pfundprb07}. However, the effects of these two interactions could not be separated. As a consequence, the exact determination of the spin-orbit mechanism was lacking.

  
In this Letter we establish the individual roles of spin-orbit and hyperfine interactions in the spin-blockade regime. Spin blockade is observed in tunable gate-defined few-electron double quantum dots in InAs nanowires. In the few-electron regime, the quantum states involved in transport can be reliably identified. This enables a careful comparison to a theory which includes random nuclear magnetic fields as well as spin-orbit mediated tunneling between triplets and singlets \cite{danonprb09}. The effects of the two interactions are traced in three distinct transport regimes, determined by the interdot coupling and the tunneling rates to the leads. The regimes are observed in two few-electron nanowire devices, 
results from one of them are discussed in the paper.

\begin{figure}[t]
\center
\includegraphics[width=8.5cm]{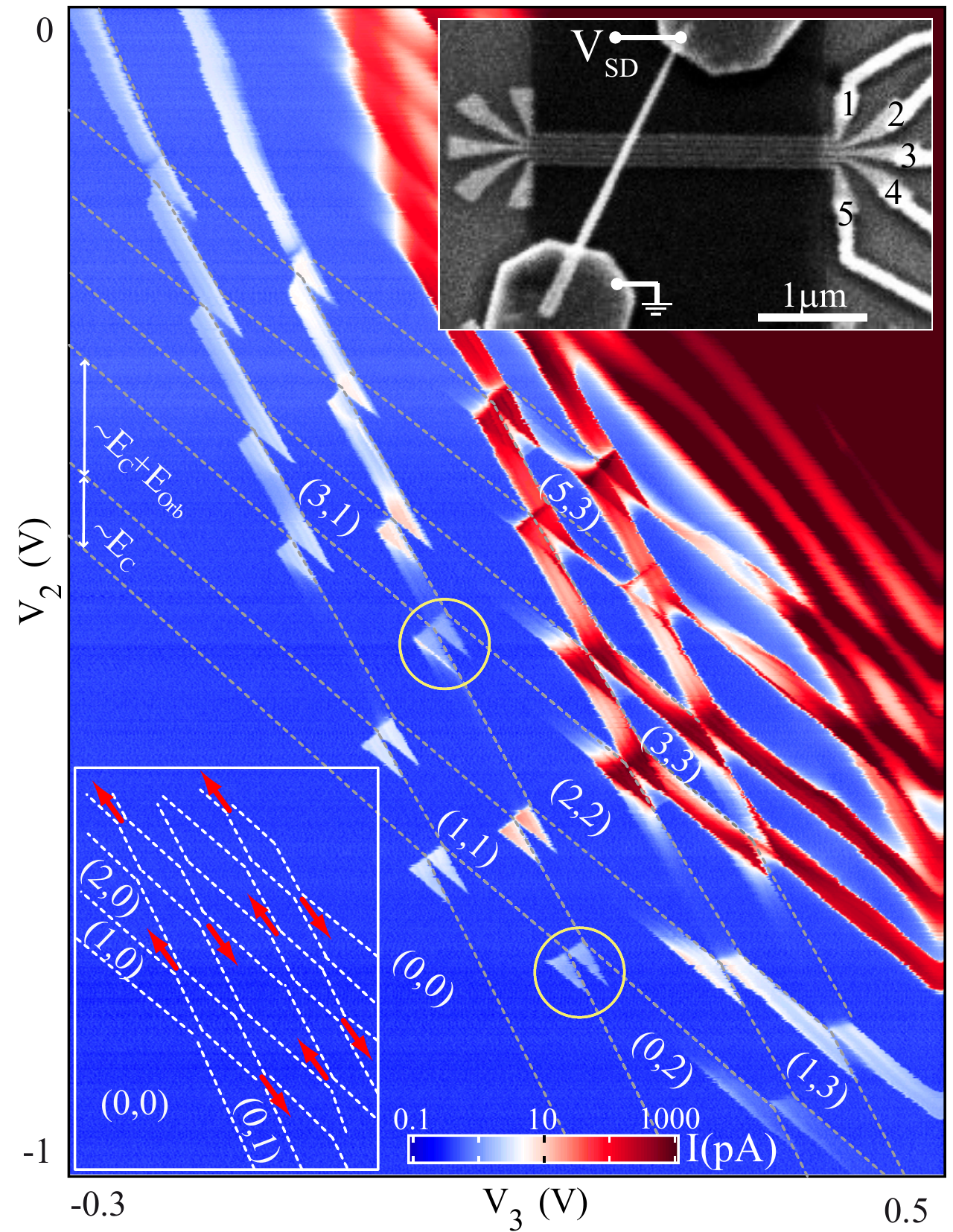}
\caption{Few-electron double dot charge stability diagram for $V_{SD}=4$~mV and $B=0$. The numbers in brackets correspond to the charges on the left and the right dots. Dashed lines separate the charge states. The energy required to add an extra electron is proportional to the spacing between the lines: $\Delta E_L=0.14e\Delta V_2$, $\Delta E_R=0.12e\Delta V_3$. The encircled regions are investigated in Fig. \ref{fig:figure2}. Upper inset: scanning electron micrograph of a nanowire device. Ti/Au gates with a pitch of $60$~nm are labeled 1-5. The black stripe is a layer of ${\rm Si_3N_4}$. Lower inset: Red arrows pointing up/down correspond to the transitions at which spin blockade is observed for positive/negative bias.}
\label{fig:figure1}
\end{figure}

The nanowire devices are fabricated on pre-patterned substrates, following Ref. \cite{fasthprl07} (Fig. \ref{fig:figure1}, upper inset). The substrates are patterned with narrow metallic gates which are covered with a 20 nm layer of Si$_3$N$_4$ dielectric to suppress gate leakage \cite{buizertprl08}. Single-crystalline InAs nanowires with diameters from 40-80 nm are deposited randomly on the substrate. Conveniently aligned wires are contacted by source and drain electrodes. Simultaneously, contacts are made to the gates underneath the wire. Measurements are performed at $T=250$~mK in a magnetic field applied perpendicular to the substrate.

The few-electron double quantum dot is formed by gates 1-4. Such tuning ensures that both dots can be emptied before the barriers become too opaque for detecting current. Gates 1 and 4 define the outer barriers, gates 2 and 3 control the interdot coupling. The charge stability diagram of a double dot is obtained by sweeping gates 2 and 3 and monitoring the source-drain current (Fig. \ref{fig:figure1}). The empty (0,0) state is verified by Coulomb blockade measurements: no lower charge states are observed in either dot up to $V_{SD}=70$~mV\cite{juriaan}. Large charging and orbital energies extracted from the last Coulomb diamond also support the few electron regime ($E_c\approx14$~meV, $E_{orb}\approx9$~meV)\cite{fasthprl07}. In both dots the energy to add the third electron ($E_c+E_{orb}$) is higher than the energy to add the second or the fourth ($E_c$), see Fig. \ref{fig:figure1}. This indicates that the first few orbitals are doubly-degenerate due to spin.

The spin states of the double dot are probed through spin blockade. A transition is spin-blocked when it is energetically allowed, but forbidden by spin conservation \cite{onoscience02}. Current can flow through a double dot via a cycle of charge states. For example the cycle $(0,1)\to(1,1)\to(0,2)\to(0,1)$ transfers one electron from left to right (Fig. {\ref{fig:figure2}(a)). The transition $(1,1)\to(0,2)$ is forbidden when the (1,1) state is a triplet and the only accessible (0,2) state is a singlet. Therefore, spin blockade suppresses the current at this charge cycle. We observe spin blockade at several charge cycles that involve ${\rm (odd,odd)} \to{\rm (even,even)}$ transitions for the first few electrons (Fig. \ref{fig:figure1}, lower inset), as expected from simple spin filling \cite{johnsonprb05}.

\begin{figure}[t]
\center \includegraphics[width=8.5cm]{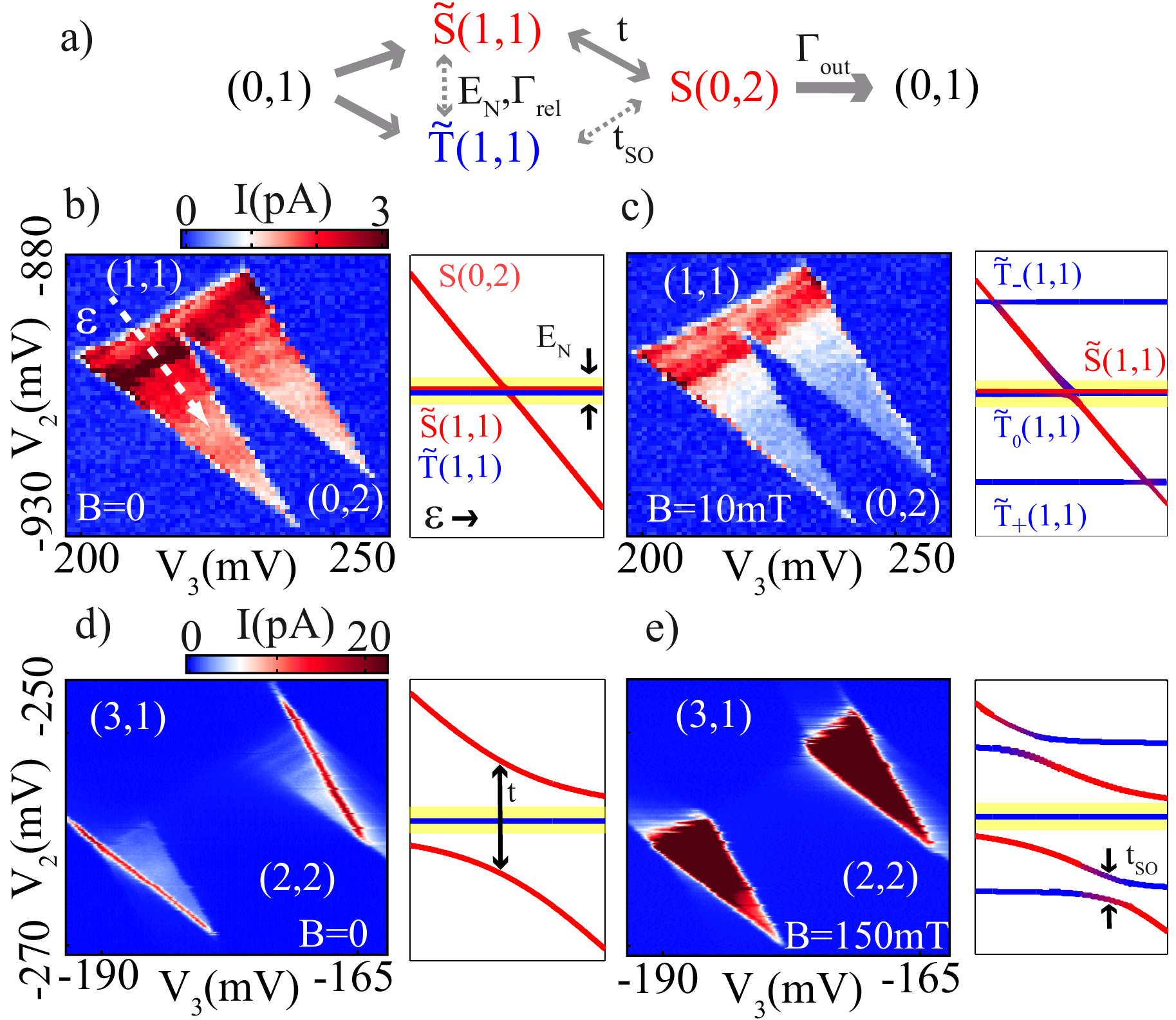}
\caption{(a) Transport diagram through a spin-blocked charge cycle at small detuning, with the relevant transition rates. (b),(c): $(1,1)\to(0,2)$ tuned to weak interdot coupling for $B=0$ and $B=10$~mT, $V_{SD}=5$~mV. (d),(e): $(3,1)\to(2,2)$ tuned to strong interdot coupling for $B=0$ and $B=150$~mT, $V_{SD}=1.3$~mV. Energy levels of (1,1) and (0,2) states are calculated for the regimes in (b)-(e). The blue-red gradient illustrates triplet/singlet hybridization. The white dash in (b) indicates a cut along the detuning axis, $\varepsilon$.}
\label{fig:figure2}
\end{figure}

An incomplete spin blockade results in finite current through the double dot. This current is due to processes that enable transitions out of triplet (1,1) states (dashes in Fig. \ref{fig:figure2}(a)). It was established in experiments on GaAs dots  that hyperfine mixing results in transitions between different (1,1) states \cite{koppensscience05, johnsonnature05, jouravlevprl06}. Ref. \cite{danonprb09} predicts that spin-orbit interaction can also lift spin blockade by hybridizing triplet (1,1) states with S(0,2). Bellow we describe how the contributions of the two interactions can be disentangled.

Flip-flops involving the fluctuating nuclear spin bath mix the (1,1) electron spin states only if they are close in energy. The characteristic energy scale over which the hyperfine interaction is effective is $E_N=AI/\sqrt{N}$ \cite{merkulovprb02}, where A is the hyperfine constant, $N$ is the number of nuclei in the dot and $I$ is the average nuclear spin.  The corresponding r.m.s. of nuclear field fluctuations is given by $B_N=E_N/g\mu_B$. (We measure the Land\'e g-factor $g=8.3\pm0.6$ by excited state spectroscopy.)

Due to spin-orbit interaction the (1,1) eigenstates become superpositions of spin triplets and the (1,1) singlet. We denote these (1,1) eigenstates with \~T$_-$, \~T$_0$, \~T$_+$ and \~S. The spin singlet admixture in \~T states is of the same order as the ratio of the dot size to the spin-orbit length $l_{dot}/l_{SO}$. Because they contain a singlet component, \~T states are coupled to S(0,2), which remains a spin singlet since both electrons in it belong to the same orbital. The exact coupling between \~T(1,1) and S(0,2) depends on the microscopic properties of the spin-orbit interaction in InAs nanowires  and on the details of confinement \cite{golovachprb08}. Here we simply parametrize this coupling with $t_{SO}\sim(l_{dot}/l_{SO})t$, where $t$ is the tunnel coupling between S(1,1) and S(0,2).


The energy levels calculated for weakly and strongly coupled double dots are shown in Figs. \ref{fig:figure2}(b)-(e) as a function of the energy detuning $\varepsilon$ between the (1,1) and (0,2) states. The calculation of the levels includes $t_{SO}$ while disregarding the effect of nuclear spins. The effect of $E_N$ is represented by a yellow stripe: the (1,1) states within the stripe are mixed by the nuclei.

The principal roles of spin-orbit and hyperfine interactions can be illustrated by tuning the interdot tunnel coupling (Fig. \ref{fig:figure2}). For small $t$,$t_{SO} \ll E_N$, the hyperfine-induced spin mixing dominates. The energy levels appear the same as for real spin singlets and triplets (Fig. \ref{fig:figure2}(b),(c)) \cite{hansonrmp07}. In this limit the current is high at zero magnetic field, but is suppressed by a small magnetic field. This occurs for fields $B\gtrsim B_N$ when the hyperfine mixing of the split-off states \~T$_+$ and \~T$_-$ with the decaying (1,1) state is reduced.

The energy levels become noticeably modified when $t_{SO}\propto t$ is large (Fig. \ref{fig:figure2}(d),(e)). But the effect of this modification can only be seen at finite magnetic field. At zero field only one of the four (1,1) states is coupled to S(0,2) by the strength $t$ (Fig. \ref{fig:figure2}(d)). The hyperfine mechanism cannot facilitate the escape from the uncoupled states because of the large singlet anticrossing, so the current is suppressed \cite{koppensscience05}.  At finite field, however, the eigenstates \~T$_+$ and \~T$_-$ are coupled to the singlet S(0,2) by a large $t_{SO}$ and the current increases (Fig. \ref{fig:figure2}(e)). The current at finite field is limited by the escape rate from the remaining one blocked state.

\begin{figure}[t]
\center
\includegraphics[width=8.5cm]{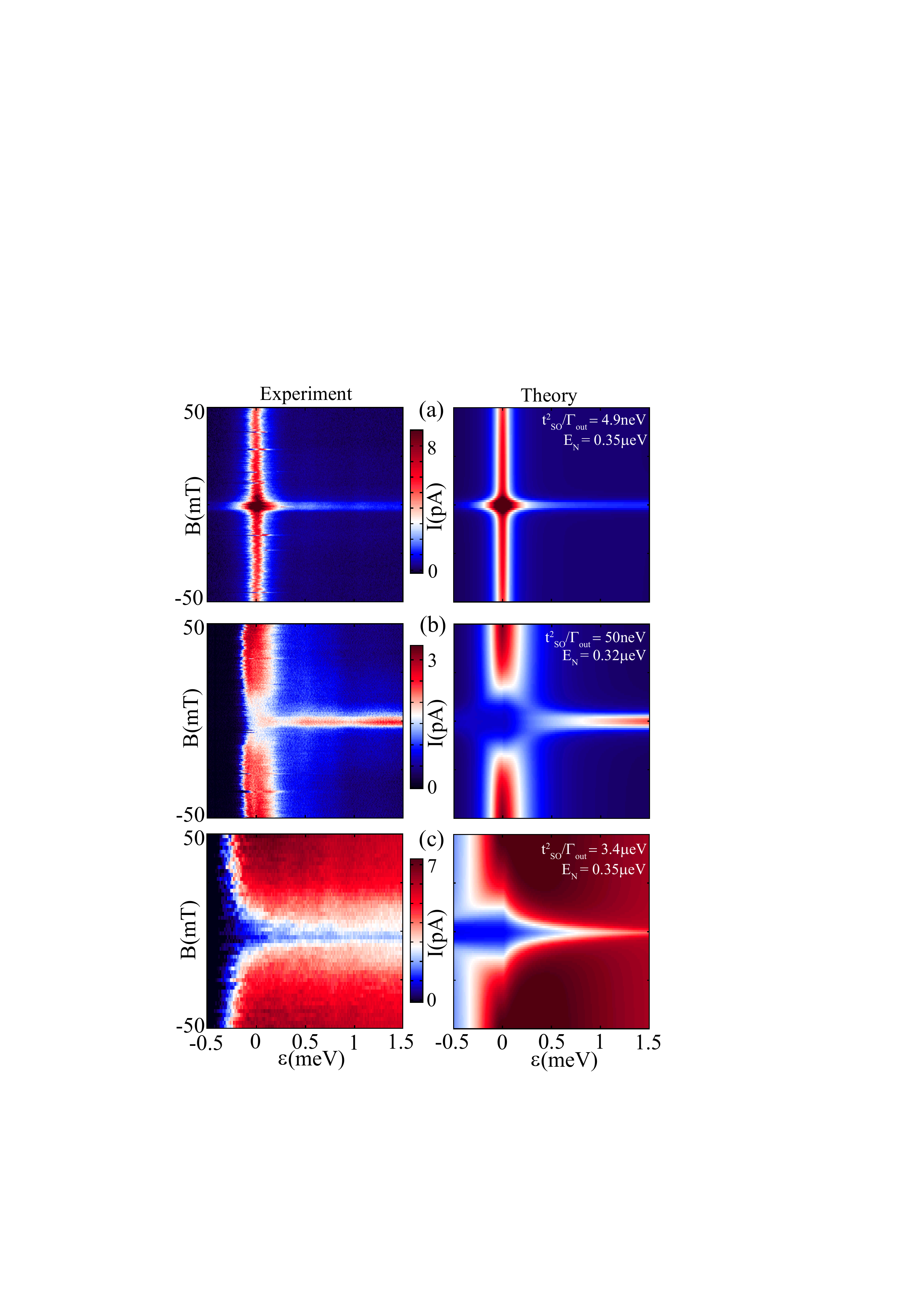}
\caption{(Left) Measured double dot current as function of detuning and magnetic field. (Right) Simulations of current for different values of $t_{SO}^2/\Gamma_{out}$ averaged over $N_f=1000$ random nuclear configurations. (a) and (b) data from $(1,1)\to(0,2)$ transition, (c) data from $(1,3)\to(2,2)$ to illustrate large $\Gamma_{rel}$ and $\Gamma_{inel}$ (see Fig. \ref{fig:figure4}). Simulation parameters: (a) $\Gamma_{out}=100$ $\mu$eV, $t=6.6$ $\mu$eV, $t_{SO}=0.7$ $\mu$eV, $\Gamma_{rel}=0$; (b) $\Gamma_{out}=70$~$\mu$eV, $t=32$~$\mu$eV, $t_{SO}=1.8$~$\mu$eV, $\Gamma_{rel}=0.2$~MHz; (c) $\Gamma_{out}=20$~$\mu$eV, $t=45$~$\mu$eV, $t_{SO}=8.2$~$\mu$eV, $\Gamma_{rel}=5.4$~MHz;}
\label{fig:figure3}
\end{figure}

In a nutshell, hyperfine interaction lifts spin blockade for weak coupling and small fields, spin-orbit interaction - for strong coupling and large fields. The current may exhibit either a hyperfine-induced peak at zero magnetic field, or a dip due to spin-orbit interaction. The interplay of the two contributions gives rise to three distinct regimes as shown in Fig. \ref{fig:figure3}. In the first regime, for weakest coupling, a zero field peak is observed for any detuning (Fig. \ref{fig:figure3}(a)). In the intermediate regime, a dip around zero detuning becomes a peak at higher detuning. For the strongest coupling, the current only shows a dip at zero field (Fig. \ref{fig:figure3}(c)).  In all regimes the high-detuning behavior extends up to $\varepsilon=5-7$~meV, where the (1,1) states are aligned with \~T(0,2) and spin blockade is lifted. The three regimes were observed at several spin-blockaded transport cycles, here we show the data from two of them (circles in Fig. \ref{fig:figure1}).

The data are in good agreement with our simple transport theory that accounts for spin-orbit and hyperfine interaction \cite{danonprb09}. The three regimes are distinguished by the rate $t_{SO}^2/\Gamma_{out}$, where $\Gamma_{out}$ is the rate of escape from the S(0,2) into the outgoing lead (in $\mu$eV). Intuitively, $t_{SO}^2/\Gamma_{out}$ is the \~T(1,1) escape rate due to $t_{SO}$. When $t_{SO}^2/\Gamma_{out}\ll E_N$  hyperfine mixing is the most effective process in lifting the spin blockade, see Fig. \ref{fig:figure2}(a). This is the case in Fig. \ref{fig:figure3}(a), where we observe a zero-field peak in the current. As $t_{SO}^2/\Gamma_{out}$ is increased, we observe intermediate regime (Fig. \ref{fig:figure3}(b)). Still, zero field peak persists at large detuning since $t^2_{SO}/\Gamma_{out}$ becomes suppressed $\propto 1/\varepsilon^2$ due to a reduced overlap of the (1,1) states with S(0,2). Around zero detuning, however, the hyperfine mixing at small fields is weaker than the spin-orbit coupling at finite fields, leading to a zero-field dip.  In the third regime, for even higher $t_{SO}^2/\Gamma_{out}\gg E_N$, the zero-field dip is extended to high positive detuning (Fig. \ref{fig:figure3}(c)). It should be stressed that the effects of both hyperfine and spin-orbit interactions are observed in all three regimes: current at higher fields is always enabled by spin-orbit interaction, and around zero magnetic field current is in part due to hyperfine mixing even for $t_{SO}^2/\Gamma_{out}>E_N$.

The peaks, dips and their widths, as well as the current levels are reproduced by a numerical simulation of transport through the spin-orbit eigenstates. The double dot current is obtained from stationary solutions of master equations \cite{danonprb09}. Spin mixing due to hyperfine interaction is included by averaging over thousends random nuclear fields. Current at high positive detuning is modeled by the inelastic transition rate, $\Gamma_{inel}=t^2 f(\varepsilon)$ from S(1,1) and $\Gamma_{inel}=t_{SO}^2 f(\varepsilon)$ from the \~T(1,1) states. The function $f(\varepsilon)$ reflects the phonon density of states in the nanowire. We determine this function by matching the inelastic current in each regime. All three regimes are reproduced with $t_{SO}=(0.12\pm0.07)t$ and $E_N=0.33\pm0.05$~$\mu$eV.  The spin-orbit length $l_{SO}\approx (t/t_{SO})l_{dot}=250\pm150$~nm can be estimated using $l_{dot}=\hbar/\sqrt{E_{orb}m_{eff}}\approx 20$~nm ($m_{eff}=0.023m_e$ in InAs). The values for $E_N$ are in agreement with the $N=10^6$ nuclei estimated from the dot size and $AI\approx350$~$\mu$eV. The values for $l_{SO}$ and $E_N$ are as expected for InAs nanowires quantum dots \cite{fasthprl07,pfundprl07}.

\begin{figure}[t]
\center
\includegraphics[width=8.5cm]{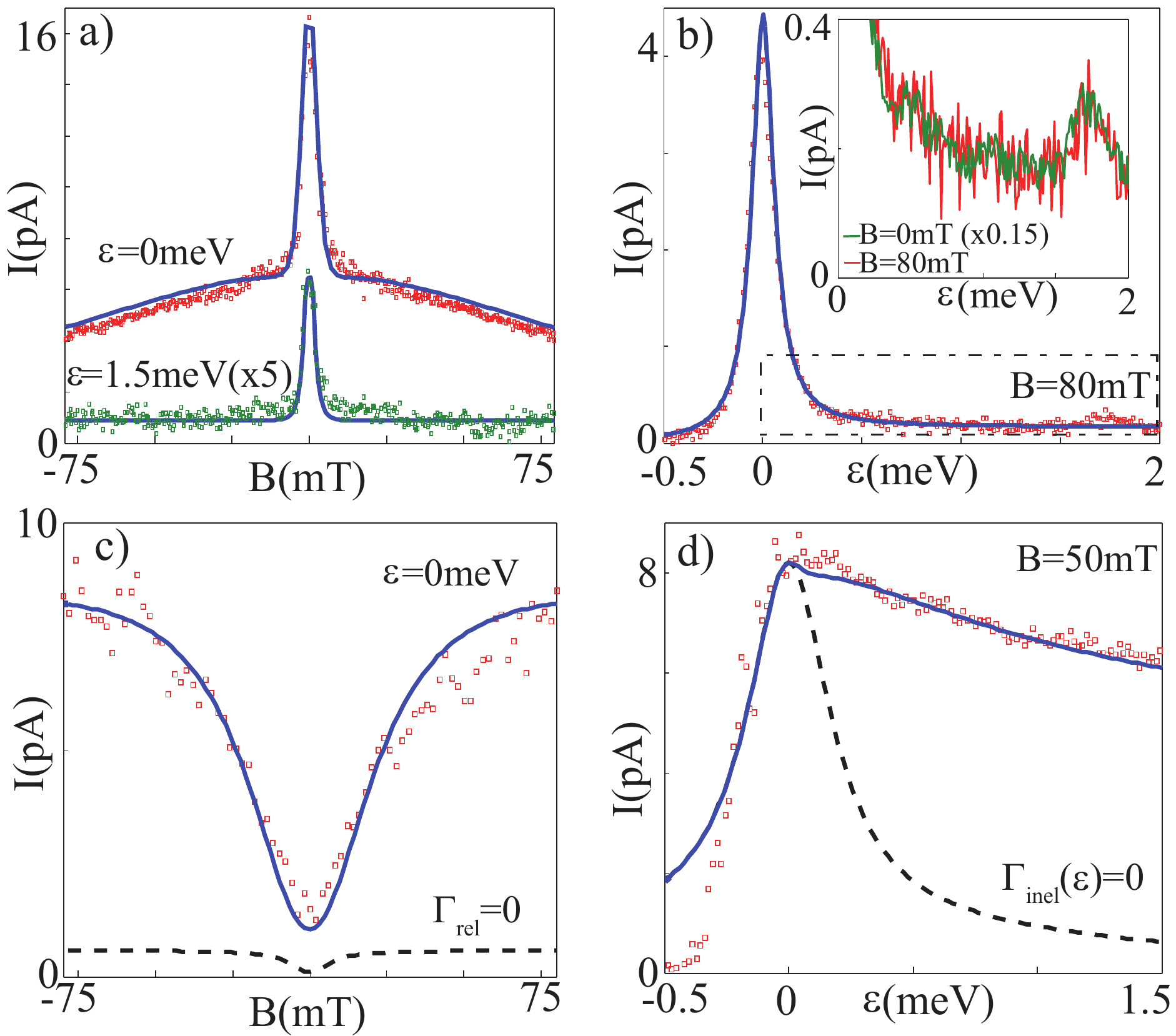}
\caption{(a),(b) Linecuts from Fig. \ref{fig:figure3}(a). The traces at $\varepsilon=1.5$~meV and $B=0$~mT are scaled by factors of 5 and 0.15 respectively. Dashed area in (b) is shown in the inset. (c),(d) Linecuts from Fig. \ref{fig:figure3}(c) and fits to the model for various $\Gamma_{rel}$ and $\Gamma_{inel}$. In the entire figure solid blue lines are simulations using parameters from Fig. \ref{fig:figure3} averaged over $N_f=30000$ (a) and $N_f=5000$ (b)-(d) random nuclear configurations. }
\label{fig:figure4}
\end{figure}

We now turn to a more quantitative analysis. The model is especially successful in reproducing the data in Fig. \ref{fig:figure3}(a), where $t_{SO}^2/\Gamma_{out}\ll E_N$. In Figures \ref{fig:figure4}(a) and \ref{fig:figure4}(b) the linecuts along magnetic field and detuning are fitted using the same set of model parameters. The model allows to trace the influences of spin-orbit and hyperfine interactions through various features of the data. The narrow peak at zero field is mainly due to hyperfine mixing (Fig. 4(a)), similar to that observed in GaAs dots \cite{koppensscience05, johnsonnature05, jouravlevprl06}. However, the wider Lorentzian background at zero detuning is due to the strong spin-orbit coupling in InAs nanowires. The elastic current drops for $B\gtrsim\Gamma_{out}/2g\mu_B\approx 100$~mT, where the detuning between \~T$_{\pm}$(1,1) exceeds the level broadening of S(0,2) set by $\Gamma_{out}$.

The current is suppressed in the inelastic regime, that is for detuning $\varepsilon\gtrsim\Gamma_{out}$ (Fig. \ref{fig:figure4}(b)). The remaining current, however, conveys information about the strength of spin-orbit interaction. At zero magnetic field the current is limited by the singlet tunneling $\sim t$, which is weak in this regime. At higher field the slowest process is the tunneling from \~T$_\pm$ states with a rate limited by $t_{SO}$, which is even weaker. The model \cite{danonprb09} predicts a simple relation $I(B=0)/I(B \gg B_N)=t^2/12t_{SO}^2$. The inset to Fig. \ref{fig:figure4}(b) shows that the current at zero field scales to the current at finite field. From the ratio we determine $t_{SO}=(0.11\pm0.02)t$ for this regime.

The model helps identify another spin relaxation mechanism present in some of the data, such as shown in Fig. 3(c) and Fig. 4(c). A zero-field dip in the elastic current is reproduced by including the hyperfine mixing and the spin-orbit hybridization. However, the predicted current is much lower than in the experiment (dashed line in Fig. 4(c)). This discrepancy can be reconciled by introducing a field-independent rate of spin relaxation $\Gamma_{rel}\approx 6$~MHz which mixes all (1,1) states \cite{danonprb09}. This spin relaxation may be induced by electron-nuclear flip-flops mediated by phonons \cite{erlingssonprb02}, spin-spin interactions mediated by charge fluctuations and spin-orbit interaction \cite{trifprb08,flindtprl06} or by virtual processes such as cotunneling or spin exchange with the leads.  The magnitude of $\Gamma_{rel}$ depends on the gate settings, and is not directly related to the magnitudes of $t$ or $\Gamma_{out}$. 

 In this regime we also observe large inelastic current (Fig. \ref{fig:figure3}(c)), which implies a high inelastic rate $\Gamma_{inel}$. Figure \ref{fig:figure4}(d) shows the contribution of inelastic current compared to the expected elastic current. Some peculiarities of the data in Figs. \ref{fig:figure3}(b), (c) are not captured by the model. The current onset is unexpectedly sharp as the detuning is increased (Figs. \ref{fig:figure3}(b), (c), \ref{fig:figure4}(d)). A possible reason for this discrepancy could be dynamic nuclear polarization not included in our model. It is known that dynamic nuclear polarizations can cause sharp current switches \cite{onoprl04}. Another explanation is that a sharp inelastic resonance at small detuning enhances the current \cite{weberprl10}.

In conclusion, we separate the effects of spin-orbit and hyperfine interactions in the spin-blockade regime of a double quantum dot. These findings will guide the development of spin-orbit controlled qubits. Further insights into spin-orbit interaction in nanowires can be obtained from direct measurements of spin coherence times.

We thank K.C. Nowack, L.M.K. Vandersypen and M.C. van der Krogt for their help. This work has been supported by NWO/FOM (Netherlands Organization for Scientific Research)  and through the DARPA program QUEST.
\bibliographystyle{apsrev}
\bibliography{sb}{}
\end{document}